\documentclass[twoside,final]{pro12}
\usepackage[ansinew]{inputenc}
\usepackage{amsmath}
\usepackage[dvips]{graphicx} 

\head{F. Mottez} {Planets around pulsars}

\begin{document}

\title{ON THE POSSIBILITY OF RADIO EMISSION OF PLANETS AROUND PULSARS.}
\author{Fabrice Mottez {\textsl LUTH -  Obs. Paris-Meudon/CNRS - Meudon -France}}


\maketitle

\begin{abstract}
A planet orbiting around a pulsar would be immersed in an ultra-relativistic under-dense plasma flow.
It would behave as a unipolar inductor, with a significant potential drop along the planet.
As for Io in Jupiter's magnetosphere, there would be two stationary Alfvén waves, the Alfvén wings (AW), attached
to the planet. 
The AW would be supported by strong electric currents, in some circumstances comparable to those of a pulsar.
It would be a cause of powerful radio waves emitted all along the AW, and highly collimated through 
relativistic aberration. 
There would be a chance to detect these radio-emissions from Earth.
The emission would be pulses as for ordinary pulsars; their occurrence would depend on the planet-star-observer angle.
These results are still preliminary, further work needs to be done.

\end{abstract}


\section{Exoplanets around pulsars}
Wolszczan and Frail [1992] and Thorsett et al. [1993] discovered two planetary systems around a pulsar,
through pulsar timing method (Doppler shift of the pulsar's period $P$). The data about the two stars, and their planetary systems are given in Tables \ref{donnees_etoiles} and \ref{donnees_planetes}.
Possibly, other planets around other pulsars remain to be discovered. In this paper, we investigate the ability of planets around pulsars to emit radio waves.

Indeed, the flux of energy available in a pulsar's environment is huge, and even comparable to 
the gravitational energy that bounds the planet to its star. The pulsars are known to spin down, and the associated dissipated power is $\dot E_{rotation} =- M_I \Omega_* \dot \Omega_* =4 \pi^2 M_I \dot P_* / P_*^3$, where $\Omega_*$ is the neutron star's rotation velocity, $P=2 \pi/\Omega_*$ the corresponding period, $\dot \Omega_* <0$ is the rotation velocity time derivative, $M_I \sim (2/5) M_* R_*^2$ is the star's moment of inertia, $M_*$ is the neutron star's mass, and $R_*$ is its radius. 
In the case of PSR 1257+12 that hosts three planets, $\dot E_{rotation}= 2. \times 10^{27}$ W. 
The gravitational energy of planet "a" is $E_G = {G M_* M_p}/{2 a }  = 4.  \times 10^{32}$  J.
Let us assume that this planet capture the whole energy emitted from the pulsar through it section
$ \pi R_P^2$, then the ratio $(R_p/a)^2 (E_G/\dot E_{rotation})  = 1,5 \times 10^6$ yr indicates that most of the gravitational energy of the planet would be dissipated in a million year. Of course, we do not expect 
that the energy capture process would be 100 \% efficient, but this comparison proves that even with a low efficiency, there may be enough power to trigger plasma acceleration around the planet, and powerful radio-waves emissions.

There exists two main families of pulsars. The "standard pulsars" have a period $P_* \sim 1$ s, and $\dot P \sim 10^{-14}-10^{-12}$. The second family, to which belong  PSR 1257+12
and PSR 1620-26 are characterized by a period $P_* \sim 10$ ms and $\dot P \sim 10^{-20}-10^{-17}$. This is the family of the "millisecond pulsars". In the following pages, we will refer to these two families of pulsars for numerical applications. 

A planet is immersed in the pulsar's magnetosphere, and the interaction between the pulsar's dissipated energy and the planet is supposed to be of electrodynamic nature. 
To understand it, we need to know in which region of the magnetosphere is the planet.
The magnetosphere of a pulsar comprises two regions bordered  by a cylinder, called the "light cylinder" that is parallel to the star's rotation axis, with a radius $r_{LC}=c/\Omega_*$, where $c$ is the light velocity. It corresponds to an absolute limit : at the distance $r_{LC}$,  a plasma co-rotating with the star would have the velocity of light. Up to a few light cylinder radii, the structure of a pulsar magnetosphere is complex. Beyond, it is a wind, blown away from the inner magnetosphere, at relativistic velocities. 

Is a planet in the inner magnetosphere or in the wind ?
Table \ref{donnees_cylindre_lumiere} shows the radii and the corresponding keplerian orbital periods for the light cylinder of  1 s and a 10 ms pulsars. We can see that a planet orbiting inside the light cylinder would have an orbital period not exceeding 3 minutes, (being extremely close to the star) while the period of the known planets exceed several days. Therefore, it is reasonable to consider that planets orbit into the pulsar wind, far from the light cylinder. For instance, the distance of planet PSR 1257+21 "a" is $a= 9.4 \times 10^4 r_{LC}$. If this planet was orbiting around a $P=1$ s pulsar, the ratio $a/r_{LC}$ would still be large, of the order of 600.

\section{Alfvén wings}

In [Mottez and Heyvaerts, 2011a], now notated [MH],  the electromagnetic interaction of a relativistic stellar wind with a planet or a smaller body
  in orbit around the star is investigated, through analytical calculations in MHD. Special relativity effects are included. 
  
 For simplicity, the theory exposed in [MH] and summarised in this section, is developed for a pulsar with a magnetic dipole axis aligned with the rotation axis.
 The wind flow is supposed to be radial ($r$ direction in spherical coordinates), and the magnetic field
 in the companion's environment is supposed to be azimuthal ($\phi$ direction). 
 In most wind's models for aligned pulsars (see [Kirk et al. 2009] for a review),  the wind is characterized by two invariants along its flow: 
 the neutron star's magnetic flux $\Psi$, and the mass flux $f$, defined as 
 \begin{eqnarray} \label{eq_f}
f&=&\gamma_0 \rho'_0 v_0^r r^2, \\
 \label{eq_psi}
\Psi&=&r^2 B_0^r,
\end{eqnarray}
where $\gamma_0$  is the wind's Lorentz factor associated to the unperturbed wind's velocity $v_0$, and $B_0^r$ is the radial magnetic field. The Lorentz factor is expected to be high, but estimates, based on various observations and models, vary over a fairly large range from $10^{1}$ to $10^{7}$ [Kirk et al. 2009].
The condition $\vec{ E}+\vec{ v} \times \vec{ B}=0$ implies that the azimuthal magnetic field is given by,
\begin{equation} \label{eq_b0phi}
B_0^\phi=B_0^r \frac{v_0^\phi -\Omega_* r}{v_0^r} \sim -\frac{B_0^r \Omega_* r}{c}.
\end{equation}
The approximation in the right hand side term is relevant at large distance ($r >> r_{LC}$), where $\Omega_* r >> c$.

The relevance of the Alfvén wings theory  is based on a comparison of the wind's velocity, and the Alfv\'en velocity. 
The Alfv\'en speed  in the wind's frame is  
  \begin{equation} \label{eq_va}
 V_A^{'-2}=c^{-2}+c_A^{'-2}=c^{-2}+ \mu_0 \rho'_0 /{B'_0}^2,
  \end{equation}  
  where $\rho'_0$ is the proper density of proper mass of the wind and
$B'_0$ is the unperturbed field observed in the wind's frame.  
The models considering a relativistic radial wind MHD model  [Arons 2004, Kirk et al. 2009]
imply an asymptotic Lorentz factor scaling as 
\begin{equation} \label{eq_gamma_infty}
\gamma_\infty \sim \sigma_0^{1/3},
\end{equation}
where $\sigma_0$ is the magnetization factor defined by
\begin{equation}
\sigma_0 = \frac{\Omega_*^2 \Psi^2 }{\mu_0 f c^3}.
\end{equation}
Then, it is shown in [MH] that the Alfvénic Mach number $M'_A= {v_{0}^r}/{V'_A}$ scales as 
$1 - 1/(2\sigma_0^{4/3})$, that is smaller than one.

\begin{figure}
\centering
 \includegraphics[width=0.6\textwidth]{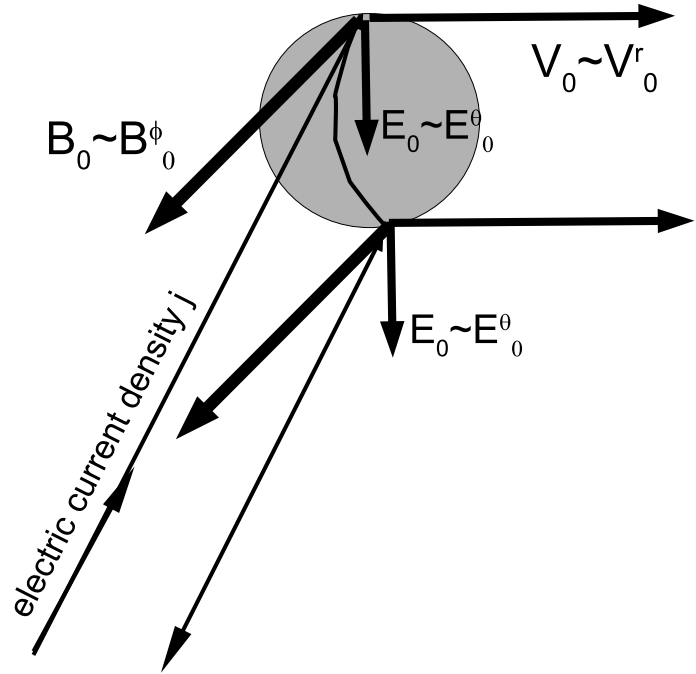}
 \caption{Schematic view of the induced electric field that generates Alfvén wings.}
 \label{fig_inducteur_unipolaire}
\end{figure}

As the fast magnetosonic waves are even faster, the wind is slower than the fast magnetosonic waves, and there is no MHD shock. Therefore, instead of being like a solar-system planet in the super-Alfvénic solar wind, the planets around a pulsar are more like Io in the sub-Alfvénic plasma flow of Jupiter's magnetosphere. 
Then, instead of being preceded by a bow shock, the pulsar's planet has a direct contact with the wind, and following the idea developed by Neubauer [1980], there is a pair of electric currents carried by  two stationary Alfv\'enic structures called Alfvén wings. They are attached to the planet on one side, and going far into the space plasma on the other side. The configuration of these electric currents is recalled in Fig. \ref{fig_inducteur_unipolaire}.
The engine of the Alfvén wings is the convection electric field $\textbf{E}_0=-\textbf{v}_0 \times  \textbf{B}_0$ induced by the motion of the wind into the magnetic field, that appears in the reference frame of the planet. It is generated by a potential drop $U$ along the body of radius $R_P$,
\begin{equation} \label{eq_unipolar_pour_application_numerique}
U = 2 R_P E_0 = \frac{2 R_P \Omega_* \Psi}{r},
\end{equation}
where $E_0= v_0^r B_0^\phi$ is directed perpendicularly to the wind flow and to the magnetic field.  


With each Alfvén wing, there is an electric circuit consisting of a incoming current carried by the Alfvén wing a coming from space, a current along the planet, and an outbound current going into space. 
The resistivity $\Sigma_{AW}$ associated to the Alfvén wing was derived from the MHD theory (see [MH]). It has a very simple form when 
the Alfvén velocity is close to $c$, as in a pulsar's wind: 
\begin{equation}
\Sigma_A \sim \frac{1}{\mu_0 c}.
\end{equation}
In that case, the order of magnitude
of the total electric current is
\begin{equation} \label{courant_total}
I_{AW}=4 (E_0 - E_i) R_P \Sigma_A,
\end{equation}
where $R_P$ is the planetary radius, and $E_i$ is an electric field along the planet caused by its ionosphere or surface finite conductivity [Neubauer 1980, MH]. 
According to Neubauer, the power dissipated by Joule effect along the ionosphere is maximized for matching of internal and external loads, that is,  when $E_i=E_0/2$.
In that case, for a $P=1$ second pulsar and an Earth-like planet at 0.2 UA, $I_{AW} \sim 10^{11}$ A. This current 
is not negligible. It has actually the same order of magnitude as the current that powers the whole pulsar magnetosphere [Goldreich and Julian 1969]. (Nevertheless, for asteroids, or for planets around recycled and less magnetized pulsars, the AW current is smaller than the Goldreich and Julian current by orders of magnitude.)

The computation of this electric current involves a first invariant $\vec{V}_s$, derived in [MH]. With $s=\pm 1$,
\begin{equation} \label{invardansR0} 
\vec{V}_s =\vec{v}-\frac{s \vec{B}(1 - {\vec{v}_0 \cdot \vec{v}}/{c^2})}{\lambda (1 -s \vec{B}_0 \cdot \vec{v}/c^2)}=\vec{v}_0-\frac{s \vec{B}_0}{\lambda  \gamma_0^2 (1 -s \vec{B}_0 \cdot \vec{v}_0/c^2)}.
\end{equation}
The second equality is an estimate of the invariant taken from the unperturbed plasma area, at large distances from the wing.
The parameter $\lambda$ can be expressed as a function of the unperturbed wind parameters,
\begin{equation}
\lambda = \left[\mu_0 \rho_0'+c^{-2} {B}_{\parallel 0}^2+ c^{-2}  \gamma_0^{-2} B_{\perp 0}^2 \right] ^{1/2}.
\end{equation}
As seen with Eq. (\ref{eq_b0phi}), at large distance from the light cylinder, $B_\phi >> B_r$ and $v_0 = v_{0 r}$. Therefore, at the planetary distances, $\vec{B}_0 \cdot \vec{v}_0/c^2 << 1$ and the invariant takes a simpler value,
\begin{equation} \label{invardansR0_simple} 
\vec{V}_s =\vec{v}_0-\frac{s \vec{B}_0}{\lambda  \gamma_0^2}.
\end{equation}
The current given in Eq. (\ref{courant_total}) is parallel to this vector.
Therefore, the angle of $\vec{V}_s$ with the radial direction determines the geometry of the Alfvén wing.
The computation of the angle $\theta$ between the wing and the radial direction is not done in [MH], and we present it in the following lines.
Considering that at the planet's distance, the magnetic field is predominantly azimuthal and $v_0 \sim c$, and $\theta$ is given by the ratio of the two perpendicular components of $\vec{V}_s$,
\begin{equation}
\theta=\arctan ({B}_0/{ c\lambda  \gamma_0^2}).
\end{equation}
The magnetic field in $\lambda$ can be expressed as a function of the invariants of the wind, $\Psi$ and $f$ given by the Eqs. (\ref{eq_f}) and (\ref{eq_psi}). In the algebraic 
development, $f$ and $\Psi$ disappear, replaced by functions of the magnetization parameter $\sigma_0$ and 
the light cylinder radius $r_{LC}$. Then,
\begin{equation} \label{eq_theta}
\theta=\arctan{ \left[\gamma_0^{-1}  \left( \frac{\sigma_0  \gamma_0^{-2} } {1+\sigma_0 \gamma_0^{-2}+\sigma_0 (r_{LC}/r)^2 } \right)^{1/2} \right]}.
\end{equation}
 
\begin{figure}
 \centering
 \includegraphics[width=0.6\textwidth]{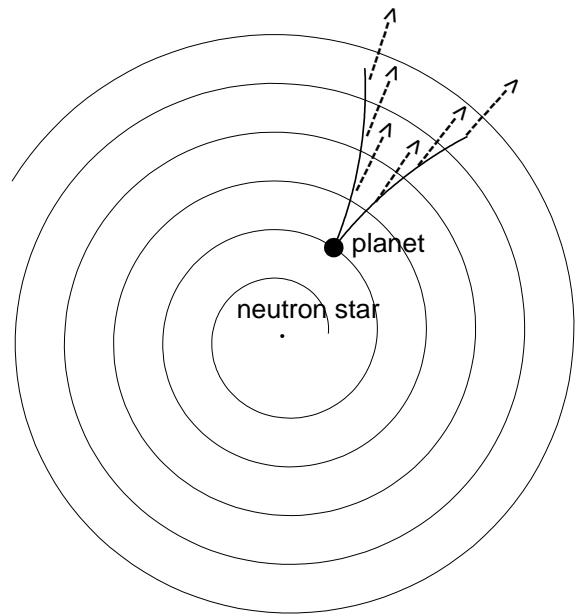}
 \caption{Sketch showing the magnetic field line passing through the companion (thin line) and the two Alfven wings (thick lines). The star (thin dot) is at the center of the figure. The companion is represented by a thicker dot. (Actually, the planet's radius is larger than that of its star.) The dashed arrows indicate the direction of propagation of the radio waves, seen from the observer's reference frame. These radiations are emitted from the Alfvén wings.}
 \label{spirale_AW}
\end{figure}

Considering the asymptotic value of the Lorentz factor given in Eq. (\ref{eq_gamma_infty}), and 
$\gamma_0 >>1$, for a planet at large distance for the star, but still in the Poynting flux dominated wind,
we find the very simple expression
\begin{equation} \label{eq_theta_infty}
\theta_\infty \sim {\gamma_\infty^{-1}} \sim \sigma_0^{-1/3}.
\end{equation}
(In this last estimate, the effect of the residual magnetic field $B_\parallel$, that is parallel to the flow, is completely neglected.)
Figure \ref{spirale_AW} shows the possible aspect of the two  Alfvén wings connected to the planet, in the pulsar wind.
The ambient magnetic field, that is spiral shaped (archimede spiral, with $B_\phi >> B_{poloidal}$) is represented by field lines (thin lines on the figure). 


\section{Radio emissions from the Alfvén wings}

We can expect that with currents as strong as in Eq. (\ref{courant_total}), the Alfvén wings associated to a planet in a pulsar's wind are strong radio-sources. 

These might be synchrotron radiation from the electrons that carry the current. But the current might be as well destabilized, and excited by transient Alfvén waves  caused by the fluctuations of the pulsar's wind passing along the planet. The conjunction of the destabilized current (filamentation, blobs with mirror points, density fluctuations) and transient Alfvén waves might cause, for instance, acceleration and the production of cyclotron maser unstable distribution functions. Such coherent radiation process might be more efficient than the sum of the synchrotron radiation from all the particles. 

We have not studied yet the associated emission mechanisms that involve highly relativistic plasmas; this topics is beyond the scope of the present report.
But we can provide an estimate of critical frequencies that would play a role in such emission processes. Let us consider the planet "c" orbiting PSR 1257+12. The magnetic field, that is mainly azimuthal, can be estimated from Eq (\ref{eq_b0phi}). The data in the tables allow to compute $\Psi \sim8.5\times 10^{12}$ Wb, and the magnetic field at the distance $a_c$ of planet "c" is $B_0(a_c) \sim 4\times 10^{-4}$ T. Therefore, the non-relativistic electron cyclotron frequency is $f_{ce} \sim 10^{7}$ Hz. The synchrotron emission spectrum of a free electron, with a perpendicular component of the velocity corresponding to a Lorentz factor $\gamma_e$ would increase up to the critical frequency $f_{crit}=f_{ce} \gamma_e^2$ [Lang, 1999]. It is expected that 
the wind's electrons and the positrons created in the star's magnetosphere  have already lost their perpendicular velocity, precisely through the synchrotron radiation. But the current could be carried, at least in parts, by particles extracted from the planet and accelerated. They could carry a non null perpendicular velocity. The spectrum of radiation of these particle would peak at $ f_{crit}$. Even for $\gamma_e \sim 100$ (that would represent a quite efficient acceleration), the spectrum would reach $f_{crit}=10^{11}$ Hz: it would remains in the radio-frequencies range (decreasing beyond millimetre wavelengths).




\section{Characteristics of the radio emissions from a planet around a pulsar}
A complete survey of the possible characteristics of the radio emissions of a planet around a pulsar remains to be done. Nevertheless, here are a few general remarks.

Whatever their emission angle in the wind's frame we can predict their cone of emission for an observer in the frame of the Earth. As the sources are embedded in a ultra-relativistic wind, the emission angles would be contained in a cone of solid angle $1/\gamma_0$ sr (relativistic aberration).

It is important to notice that the emission would not come from the planet itself, but from the Alfvén wings extending all along the planetary orbit.

If we take into account the inclination of the star's magnetic field over its rotation axis, the magnetic field seen in the planet's orbital plane varies with the star's rotation angle. Therefore, the inclination of the wind's magnetic field angle with the orbital plane oscillate. As the Alfvén wind makes a constant angle with the magnetic field, the Alfvén wind inclination over the orbital plane oscillates too. The associated wavelength is $\lambda = 2 \pi V_A/\Omega_* \sim 2 \pi  c/ \Omega_*$. Because of this oscillation, an observer would not see always the AW's radio emissions, they would appear pulsed. The observed period would be 
\begin{equation} \label{periode_apparente}
\Omega_{observed}=\Omega_* \pm \Omega_{orb},
\end{equation}
where $\Omega_{orb}$ is the planet's orbital period ($\Omega_{orb}<<\Omega_*$).



Another cause of variation of the period would be the Doppler shift associated to the varying distance of the sources (that are in the Alfvén Wing) from the planet and from us. For emissions confined in the radial direction, this Doppler shift would be a  function of the planet-star-observer angle. 

Because of the finite latitudinal extension of the AW (see Eqs. (\ref{eq_theta}) and  (\ref{eq_theta_infty}))
and because the radio-waves would be emitted in the radial direction, the possibility of their observation 
would depend on the observer-star-planet angle. As the latitudinal extension diminishes when  
the wind's Lorentz factor $\gamma_0$ increases, and because  $\gamma_0$ is expected to be high,
it is possible that the planetary radio emission are observed only during a small fraction of each orbit. The short laps of time of observability would be repeated regularly with the planet's orbital period.  
 

\section{Conclusion}
A planet embedded in the highly relativistic, highly magnetized and sub-Alfvénic wind of a pulsar generates two pairs of electric current going inward and outward the planet. Such structures have been already observed in the case of Io, where the current flows between Io and Jupiter. These ribbons of current are the electric counterparts of stationary Alfvén waves emitted by the planet. They are called Alfvén wings. In the case of a planet in a pulsar's wind, the current is connected on one side to the planet, and fades into space under the control of the Alfvén wave resistivity, that is of the order of the vacuum resistivity. 

The Alfvén wing current amplitude has been estimated in [MH]. It is comparable to that of a pulsar (the Goldreich-Julian current) for a planet around a standard "1 second" pulsar. It would be lower by three orders of magnitude in the case of recycled pulsars. Such systems of currents might be strong radio-waves emitter, especially for planets around young pulsars. The two Alfvén wings would behave as two long antennas connected on one side to the planet. 

Because of the relativistic aberration, in the frame of an observer on Earth, the Alfvén wings would emit radio waves mostly radially, in the outward direction (from the star to space). The emission angles would not describe the shape of a beam, but a fragment of a corona, beamed in the latitudinal direction, and in the azimuthal direction. The extent of the emission angle ($\theta$ in the latitudinal direction) would depend strongly on the wind's Lorentz factor $\gamma_0$ .  

Because of the (most expected) inclination of the pulsar's magnetic field over its rotation axis, the orientation of the magnetic field seen by the planet in the wind would be modulated, with a frequency close to the neutron star's rotation period. As the orientation of the Alfvén wings is determined by those of the magnetic field, their direction, and the direction of emission of the radio waves would wobble. This would result, for an observer in a fixed direction, into a pulsed emission. 

This study is preliminary. The mechanisms of the radio emission have not been studied, and a frequency spectrum is not proposed. We should just mention that this region is much less magnetized that the pulsar's inner magnetosphere where the "normal" radio emission are supposed to come from. Nevertheless, the physics of the radio emissions could not be transposed directly from Solar-system studies, because of the highly relativistic character of the pulsar's wind in which the sources would be embedded.



\begin{table}
\caption{Data infered about the two stars known to host planets.}
\label{donnees_etoiles}
\begin{center}
\begin{tabular}{|c| c| c| c| c |c|} 
   \hline 
   Name & $P$ (s) & $\Omega_*$ ($s^{-1}$) & $B_*$ (Gauss) & $M_*$ ($M_{\odot}$)& $R_*$ (km) \\ 
     & Star's rotation &$\Omega_*= \frac{2 \pi}{P}$  & Magnetic & Star's mass & Star's radius \\ 
   \hline 
   PSR 1257+12 & 0.006 & 1010. & $8.8 \times 10^{8}$ & 1.4 & $\sim 10.$  \\
   \hline 
   PSR 1620-26 &  0.011  & 567  & $3. \times 10^{9}$  & 1.35 & $\sim 10.$    \\
   \hline 
\end{tabular}
\end{center}
\end{table}

\begin{table}
\caption{Data about the planets orbiting around pulsars.}
\label{donnees_planetes}
\begin{center}
\begin{tabular}{|l |c |c |c |c |c|} 
   \hline 
   Name & $M_P$ ($M_{\oplus}$) & $R_P$ ($R_{\oplus}$) & $P_{orb}$ (day) & $a$ (AU) & $e$\\ 
   \hline 
   PSR 1257+12 a & 0.02 & 0.28 & 25. & 0.19 & 0\\
   PSR 1257+12 b & 4.3  & 1.68 & 66. & 0.36 & 0.0186\\
   PSR 1257+12 c & 3.9  & 1.62 & 98. & 0.46 & 0.0252\\
   \hline 
   PSR 1620-26 a &  794  & 9,5  & 36367.  & 23. &   \\
   \hline 
\end{tabular}
\end{center}
\end{table}

\begin{table}
\caption{Radii and corresponding keplerian orbital periods for the light cylinder of  1 s and a 10 ms pulsars.}
\label{donnees_cylindre_lumiere}
\begin{center}
\begin{tabular}{|l |c |c |c|} 
   \hline 
   Name & $P$ (s) & $r_{LC}$ (km) & $P_{orb}$ (s) \\ 
   \hline 
   Standard PSR      & 1.     & 47 000  &  147 \\
   Fast PSR & 0.010  & 477  & 0.147  \\
   \hline 
\end{tabular}
\end{center}
\end{table}

 
\section*{References}
\everypar={\hangindent=1truecm \hangafter=1}

{{Arons}, J.},
{Theory of pulsar winds},
\textit{Advances in Space Research},
\textbf{33},{466-474},2004.

N.~{Bucciantini}, T.~A. {Thompson}, J.~{Arons}, E.~{Quataert}, and L.~{Del
  Zanna},
  {Relativistic magnetohydrodynamics winds from rotating neutron
  stars}.
\textit{Monthly Notices of the Royal Astronomical Society},
  \textbf{368},1717--1734, 2006.

T.~{Chust}, A.~{Roux}, W.~S. {Kurth}, D.~A. {Gurnett}, M.~G. {Kivelson}, and
  K.~K. {Khurana},
 {Are Io's Alfv{\'e}n wings filamented? Galileo observations}.
\textit{Planetary and Space Science}, \textbf{53},395--412, 2005.

P.~{Goldreich} and W.~H. {Julian},
 {Pulsar Electrodynamics}.
\textit{Astrophysical Journal}, \textbf{157},869--+, 1969.

S.~{Hess}, F.~{Mottez}, and P.~{Zarka},
 {Jovian S burst generation by Alfv{\'e}n waves}.
\textit{Journal of Geophysical Research (Space Physics)},
  \textbf{112(11)},11212--+, 2007b.

S.~{Hess}, F.~{Mottez}, and P.~{Zarka},
 {Effect of electric potential structures on Jovian S-burst
  morphology}.
\textit{Geophysical Research Letters}, \textbf{36},14101--+, 2009b.

S.~{Hess}, P.~{Zarka}, and F.~{Mottez},
 {Io-Jupiter interaction, millisecond bursts and field-aligned
  potentials}.
\textit{Planetary and Space Science}, \textbf{55},89--99, 2007.

S.~{Hess}, P.~{Zarka}, F.~{Mottez}, and V.~B. {Ryabov},
{Electric potential jumps in the Io-Jupiter flux tube}.
\textit{Planetary and Space Science}, \textbf{57}, 23--33,  2009.

S.~L.~G. {Hess}, P.~{Delamere}, V.~{Dols}, B.~{Bonfond}, and D.~{Swift},
{Power transmission and particle acceleration along the Io flux
  tube},
\textit{Journal of Geophysical Research (Space Physics)},
  \textbf{115(14)}:6205--+, 2010.

{{Kirk}, J.~G. and {Lyubarsky}, Y. and {Petri}, J.},
{The Theory of Pulsar Winds and Nebulae},
\textit{Astrophysics and Space Science Library},
  \textbf{357}:421-+, 2009.
  
K.R. Lang,
Astrophysical Formulae,
\textit{A \& A Library, Springer}, third edition, 1999.

F.~C. {Michel}, {Relativistic Stellar-Wind Torques}.
\textit{Astrophysical Journal}, \textbf{158},727--+, 1969.

[MH] F.~{Mottez} and J.~{Heyvaerts},
   {Magnetic coupling of planets and small bodies with a pulsar wind.}
\textit{submitted to Astronomy and Astrophysics}, 2011a.

F.~{Mottez} and J.~{Heyvaerts},
   {A magnetic thrust action on the orbit of small bodies around a pulsar.}
\textit{submitted to Astronomy and Astrophysics}, 2011b.

F.~M. {Neubauer}, {Nonlinear standing Alfven wave current system at Io - Theory}.
\textit{Journal of Geophysical Research (Space Physics)}, \textbf{85},1171--1178, 1980.

J.~{Queinnec} and P.~{Zarka}, {Io-controlled decameter arcs and Io-Jupiter interaction}.
\textit{Journal of Geophysical Research (Space Physics)},
  \textbf{103},26649--26666, 1998.

S.~E. {Thorsett}, Z.~{Arzoumanian}, and J.~H. {Taylor},
{PSR B1620-26 - A binary radio pulsar with a planetary companion?}
\textit{Astrophysical Journal Letters}, \textbf{412}, L33--L36, 1993.

A.~{Wolszczan} and D.~A. {Frail},{A planetary system around the millisecond pulsar PSR1257 + 12}.
\textit{Nature}, \textbf{355}:145--147, 1992.


\end{document}